\documentclass[aps,prb,twocolumn]{revtex4} 

\usepackage{graphicx}
\usepackage{dcolumn}
\usepackage{bm}

\newcommand{\comment}[1]{}

\begin{document}

\title{Herding Shr\"odinger's  Suppos\'ed Cat.
How the Classical World Emerges from the Quantum Universe.}


\author{Phil Attard}
\affiliation{{phil.attard1@gmail.com}}


\begin{abstract}
The pointer states and preferred basis of the classical world
are that of definite positions and momenta.
Here it is shown that the reason for the absence of superposition states
is the limited resolution with which observations or measurements are made.
\end{abstract}


\maketitle

%
\section{Introduction}
\setcounter{equation}{0} \setcounter{subsubsection}{0}
%

``Entities are not to be multiplied without necessity''\\
---Law of Parsimony (William of Occam, 1287--1347).\cite{Occam}

There are four main attributes that distinguish the quantum world
from the classical:
first, the quantization of states;
second, the uncertainty in simultaneous values of position and momentum;
third, occupancy of particle states consequent on wave function symmetrization;
and fourth, the superposition of states.

The first two of these are quantified by Planck's constant,
and unless a measurement or phenomenon
is contingent on scales of this size or smaller
then it will appear to be classical.
These two types of quantum phenomena are negligible
when the temperature is high enough.
\cite{QSM,STD2}
The third attribute also depends on the value of Planck's constant,
as it turns out that symmetrization effects may be neglected
when the relevant particle separations
are larger than the thermal wave-length,
which is proportional to Planck's constant.
In practice this means that this type of quantum phenomenon
is negligible for high enough temperatures
and low enough densities.
\cite{QSM,STD2}

The superposition of states is in a different category
to the other three quantum attributes
as it does not appear to directly depend upon the value of Planck's constant
(but see the text).
Quantum mechanics, specifically Schr\"odinger's equation,
is a linear theory.
Hence the sum of valid wave functions,
including macroscopically distinct wave functions,
is also a valid wave function.
But in the classical universe such linear superpositions of observed states
are not themselves observed.

Schr\"odinger's cat embodies
the divide between the quantum and the classical universe
manifest in the superposition of states.
Hidden from view,
the cat is said to be simultaneously alive and dead,
in accordance with the rules of quantum mechanics.
But upon observation the cat is exclusively alive or dead,
which manifests the prohibition on superposed states
in the classical world.
Although the common-sense view is that the cat is metaphorical,
a significant number of otherwise intelligent scientists
appear to take the cat quite literally.
How is one to reconcile these qualitatively different pictures?
How exactly does classical reality
emerge from the underlying quantum equations?

%
\section{Analysis}
\setcounter{equation}{0} \setcounter{subsubsection}{0}
%


\subsection{What is Superposition?}

Hilbert space
is closed under scalar multiplication and addition,
$\Psi = \lambda_1 \Psi_1 + \lambda_2 \Psi_2$, for example.
In consequence
any wave function can be written as a series of orthonormal basis functions,
\begin{equation}
|\psi\rangle = \sum_n \psi_n |n\rangle
, \;\; \psi_n = \langle n | \psi \rangle.
\end{equation}
In so far as $n$  labels a state of the system,
this expression suggests that the system in the  wave state $\psi$
is actually in a superposition of states $n$.

Although the use of the words `state' and `superposition'
have an arguable justification in this context,
they don't actually contribute to resolving the present question
of the absence of classical superposition.
This is because a wave function cannot be measured or observed directly.

Instead it is expectation values that correspond
to the physical act of observation or measurement.
More precisely, they predict the outcome of a future measurement.
The expectation value of an arbitrary operator $\hat A$
when the system is in the normalized wave state $\psi$ is
\begin{eqnarray}
\langle \psi| \hat A | \psi \rangle
 & = &
\sum_{n,m} \psi_n^* \psi_m A_{nm}
 \\ & = &
\sum_{n} \psi_n^* \psi_n A_{nn}
+
\sum_{n,m}\!^{(n\ne m)} \psi_n^* \psi_m A_{nm} .\nonumber
\end{eqnarray}
The second equality separates the diagonal
and the off-diagonal contributions.

For the diagonal terms the quantity $A_{nn}$ has the interpretation
of the value of the operator in the state $n$.
Since $\psi_n^* \psi_n = |\psi_n |^2$ is real and non-negative,
it has the interpretation of the probability or weight of the state $n$.
Probability and weight here mean the same as in classical probability theory.
\cite{TDSM}

The off-diagonal contributions $ n \ne m$ are more difficult to interpret
in terms of classical concepts.
One could say that $A_{nm}$ represents the coupling of states $n$ and $m$
for the operator $\hat A$,
and that $\psi_n^* \psi_m$, which is complex,
represents the amount of interference between the states $n$ and $m$
in the current wave state $\psi$.
This interference reinforces or cancels the coupling.

I argue that the phrase `superposition of states'
is appropriate
for these off-diagonal contributions.
The physical interpretation of pairwise coupling and interference
is not inconsistent with the physical picture of the system
being simultaneously in all the states $n$.
Alternatively,
one might interpret it as multiple systems, each in its own state,
being superposed on each other.
Of course this simultaneous occupancy of the states
is hard to visualize classically.
Ultimately
the physical picture one envisages makes no actual difference
since the only thing that is essential is the mathematical manipulation
of the wave functions, operators, and expectation value.
It would not matter if one simply considers the off-diagonal terms
as non-classical probabilities whose meaning is to be garnered
from their mathematical manipulation.
But in either case it is important that the words used reflect and define
the underlying concept,
and in a real sense the word `superposition'
does convey the idea of the mutual interference of the states
in the off-diagonal contribution.

I further argue that the word `superposition'
ought to be reserved for the off-diagonal terms alone.
One should not use it to describe the diagonal contribution
because this has the clear interpretation as classical probability.
In this theory, giving the probabilities for the possible outcomes
of an event is most definitely not a statement that all outcomes
occur simultaneously.
For example, assigning probabilities to horses in a race
does not mean that all the horses will simultaneously win the race.
Similarly calculating the diagonal terms in the expectation value
is a prediction of the outcome of a future measurement,
not a statement that the system will be measured
to be simultaneously in all the states.
Classical probability can be understood as merely the mathematical
representation of physical weight,
which are necessarily real and non-negative.\cite{TDSM}
Some understand it instead as a measure of information or belief.
For our purposes all that matters is that classical probability
is restricted to systems in which the occupancy of the states
is mutually exclusive.\cite{TDSM}
Obviously this mutual exclusivity
is the opposite of the normal understanding of the word `superposition'.
For this reason
it is inappropriate to use the latter
for the diagonal contribution to the expectation value.

Whether or not one agrees with this definition
of the superposition of states,
or even if one regards it as overly pedantic,
there should be no quibble with the assertion that
the diagonal contribution to the expectation value has a classical character,
and the non-diagonal contribution is non-classical.
And so the question of why there is no superposition of states
in the classical universe boils down to
the mathematical question of
why classically only the diagonal terms contribute to expectation values.

\subsection{Classical Measurement}

Perhaps surprisingly,
like the other three quantum attributes mentioned above,
the key to the answer also lies in the value of  Planck's constant.
We shall take it as axiomatic that our observation and measurement
of the classical world is limited to a resolution
that is large on the scale of Planck's constant.
Of course in reality it is possible to actually measure quantum effects,
and so there are some measurement techniques with higher resolution than this.
Also, there exist  quantum phenomena that are macroscopic in character.
But our concern here is the absence of superposition
in classical systems,
and for our purposes we shall define such systems to be those
that are measured or observed with limited resolution.
The following analysis of the effects of limited resolution
on the superposition of states
says nothing about other quantum phenomena.

This definition of a classical measurement
as one of limited resolution
does not \emph{a priori} preclude
the superposition of states in classical systems.
One has to actually show how the laws of quantum mechanics
determine that the off-diagonal contribution
to an expectation value  vanishes in such systems.

It will be shown this definition of a classical system
provides a  quantitative condition
sufficient for the absence of the superposition of states.
This establishes a useful criterion to guide
thinking and discussion of the physical universe,
and it gives the reason for the absence of superposition
in this type of classical system.
For example, a cat is a macroscopic object
the observation of which is large on the scale of Planck's constant.
Its enlightening to give explicitly the reason
why Schr\"odinger's cat can't be both dead and alive.

The answer to the question
of why and when only the diagonal terms contribute to expectation values
begins by defining  a set of classical states.
This set is not complete in Hilbert space;
the subset defined by its span may be called `the' classical space.
The classical wave functions that comprise the set
will turn out to be locally orthogonal (ie.\ no overlap)
for resolutions not less than about Planck's constant.

As a motivation for the particular set to be chosen,
consider the fact that our brains have evolved to perceive objects,
specifically their location and movement.
Hence the position and momentum of particles
play a key role in the observation and measurement of
the classical universe.
Indeed since first Newton and then Boltzmann
it has been known that the state of a classical system
can be described mathematically
by the positions and momenta of its constituent particles.
Accordingly,
the aim here is to show that there can be no superposition
of particle states when their positions and momenta
are observed or measured on scales larger than Planck's constant.

Consider the set of wave packets for a sub-system of $N$ particles,
$\phi_{{\bf p}{\bf q}}({\bf r})
=
\prod_{j=1}^N
\phi^{(1)}_{{\bf p}_j{\bf q}_j}({\bf r}_j)$.
Here ${\bf r} = \{ {\bf r}_1, {\bf r}_2, \ldots , {\bf r}_N \}$,
with ${\bf r}_j = \{ x_j, y_j, \ldots, d_j\}$
being the representation position in $d$ dimensional space.
Suppose that the wave packets have width ${\cal O}(\hbar)$
in momentum and position space.
Suppose further
that the momenta ${\bf p}$ and positions ${\bf q}$ are discrete
with sufficient separation $\Delta_p$ and $\Delta_q$ between the grid points
so that the wave packets do not overlap.
The grid points  may be called classical phase points.

The wave packet $\phi_{{\bf p}{\bf q}}$ makes a low pass filter
for changes in position and in momentum.
In this sense it provides a complete basis for measurements
with no better than classical resolution.
We shall assume that any classical phase function
is slowly varying on the scale of $\Delta_p$ and $\Delta_q$.

As an explicit example, the minimum uncertainty wave packet
for $N$ particles in $d$ dimensions is\cite{Messiah61,Merzbacher70}
\begin{equation}
\phi_{{\bf p}{\bf q}}({\bf r})
=
(2\pi\xi^2)^{-dN/4}
e^{({\bf r}-{\bf q})^2/4\xi^2}
e^{-{\bf p}\cdot ({\bf r}-{\bf q})/i\hbar} .
\end{equation}
It is straightforward to show that
these wave packets are orthonormal,
\begin{eqnarray}
\lefteqn{
\langle \phi_{{\bf p}'{\bf q}'}
| \phi_{{\bf p}''{\bf q}''}\rangle
} \nonumber \\
& = &
e^{-({\bf q}'-{\bf q}'')^2 /8\xi^2 }
e^{-({\bf p}'-{\bf p}'')^2 \xi^2/2\hbar^2 }
e^{({\bf q}''\cdot {\bf p}' - {\bf q}'\cdot {\bf p}'')/i\hbar}
\nonumber \\ & = &
\delta_{{\bf q}',{\bf q}''}\,\delta_{{\bf p}',{\bf p}''}
, \;\;
\Delta_q \gg 2 \xi , \; \Delta_p \gg \hbar/ \xi .
\end{eqnarray}

The expectation values are
\begin{equation}
\langle \phi_{{\bf p}{\bf q}}| \hat {\bf q} | \phi_{{\bf p}{\bf q}} \rangle
= {\bf q}
, \mbox{ and }
\langle \phi_{{\bf p}{\bf q}}| \hat {\bf p} | \phi_{{\bf p}{\bf q}} \rangle
= {\bf p}.
\end{equation}
The grid spacings should be chosen to be very much greater than
the root mean square uncertainty in these expectation values,
\begin{equation}
\sqrt{ \langle \phi_{{\bf p}{\bf q}}|
(\hat {\bf q} - {\bf q})^2
| \phi_{{\bf p}{\bf q}} \rangle }
= \xi \ll \Delta_q,
\end{equation}
 and
\begin{equation}
\sqrt{ \langle \phi_{{\bf p}{\bf q}}|
(\hat {\bf p} - {\bf p})^2
| \phi_{{\bf p}{\bf q}} \rangle }
= \frac{\hbar}{2\xi} \ll \Delta_p.
\end{equation}
Since these fluctuations are negligible
on the scale of the classical resolution,
the wave packet is a simultaneous position and momentum eigenfunction,
\begin{equation}
\hat {\bf q} | \phi_{{\bf p}{\bf q}} \rangle
= {\bf q} | \phi_{{\bf p}{\bf q}} \rangle
+ o(\Delta_q,\Delta_p) ,
\end{equation}
and
\begin{equation}
\hat {\bf p} | \phi_{{\bf p}{\bf q}} \rangle
= {\bf p}  | \phi_{{\bf p}{\bf q}} \rangle
+  o(\Delta_q,\Delta_p) ,
\end{equation}
as can be confirmed by direct calculation.
The fact that at this level of approximation
the wave function is a simultaneous eigenfunction
means that the position and momentum operators commute,
which is consistent with Planck's constant being negligible
for measurements made with classical resolution.

It is worth mentioning that this result stems
from the fact that the wave packet is relatively sharply peaked
in both momentum and position space.
This means that the grid spacing can be chosen
so that distinct wave packets do not overlap,
\begin{eqnarray}
 \phi_{{\bf p}'{\bf q}'}({\bf r})^*  \phi_{{\bf p}''{\bf q}''}({\bf r})
& \approx &
| \phi_{{\bf p}'{\bf q}'}({\bf r})|^2
\delta_{{\bf q}',{\bf q}''}\,\delta_{{\bf p}',{\bf p}''} ,
\nonumber \\ && 
\Delta_q \gg \xi , \; \Delta_p \gg \hbar/\xi  .
\end{eqnarray}
This result holds not only for positions but also for momenta
because of the rapid and canceling behavior that occurs
if ${\bf p}' \ne {\bf p}''$.
An analogous conclusion is reached if the argument is made
in the momentum representation.
One can therefore consider the set of wave packets
to be locally orthogonal on the coarse grained grid,
which provides a way to picture
the absence of the superposition of classical states
that is now derived.

An arbitrary normalized wave function
can be approximated as the superposition of classical phase points,
\begin{equation}
\psi({\bf r}) \approx
\sum_{ {\bf p},{\bf q} }
\psi_{{\bf p}{\bf q}} \, \phi_{{\bf p}{\bf q}}({\bf r})
, \;\; \psi_{{\bf p}{\bf q}}  = \langle \phi_{{\bf p}{\bf q}}| \psi  \rangle .
\end{equation}
This effectively filters out the high frequency parts of the wave function
by projecting the latter onto the classical space.

It is assumed that all classical phase functions
are slowly varying on the scale of the grid spacing.
Hence the expectation value of an operator
that is an arbitrary function of the momentum and position operators is
\begin{eqnarray}
\langle \psi | A(\hat{\bf p},\hat{\bf q}) |\psi \rangle
& = &
\sum_{ {\bf p},{\bf q} }
\sum_{ {\bf p}',{\bf q}' }
\psi_{{\bf p}{\bf q}}^* \, \psi_{{\bf p}'{\bf q}'}
\langle  \phi_{{\bf p}{\bf q}} | A(\hat{\bf p},\hat{\bf q}) |
 \phi_{{\bf p}'{\bf q}'} \rangle
\nonumber \\ & = &
\sum_{ {\bf p},{\bf q} }
\sum_{ {\bf p}',{\bf q}' }
\psi_{{\bf p}{\bf q}}^* \, \psi_{{\bf p}'{\bf q}'}
\langle  \phi_{{\bf p}{\bf q}} |
 \phi_{{\bf p}'{\bf q}'} \rangle
\nonumber \\ &  & \mbox{ } \times
 \left[A({\bf p}',{\bf q}') + o(\Delta_p,\Delta_q) \right]
\nonumber \\ & = &
\sum_{ {\bf p},{\bf q} }
\psi_{{\bf p}{\bf q}}^* \, \psi_{{\bf p}{\bf q}} \,
A({\bf p},{\bf q}) .
\end{eqnarray}
One sees that
only the diagonal terms contribute to the expectation value,
which is to say that there is no interference between the superposed states.
Therefore the occupancy of the position and momentum states
may be regarded probabilistically
rather than physically.
The expectation value
has the interpretation that the system is in a mixture of pure states,
with $| \psi_{{\bf p},{\bf q}} |^2$
being the  state probability.

One concludes from this that
since the quantum mechanical expectation value
is a prediction of the outcome of a measurement or observation,
than an  expectation value made with classical resolution
corresponds to the classical statistical average of possible measured values.
Since classical probability is necessarily a prediction about a system
in which  the occupancy of the possible states is mutually exclusive,
\cite{TDSM}
this formula for the predicted outcome of a classical measurement
implies a system in which superposition or interference effects are prohibited.
This probabilistic formulation is obviously consistent
with our experience of the classical world,
namely one in which the only outcomes ever observed
are absent the superposition of states.

%
\section{Discussion}
%

\subsubsection{Summary}

This paper has sought to explain the absence of the superposition of states
in the classical world by attributing it to the relatively coarse
resolution with which classical measurements or observations are made.
Although it is well-known that quantum mechanics applies
to the very small and classical mechanics to the very large,
this in and of itself
does not explain the absence of classical superposition.
Arbitrary wave functions can be validly superposed in quantum mechanics,
including those that differ macroscopically.
Hence there is a puzzle to be resolved here.

The present analysis shows that  observations or measurements
made with limited precision
act like a low-pass filter that suppresses rapid variations
in the wave function.
In turn this induces a local orthogonality
that effectively makes any operator block diagonal.
An expectation value thus becomes the equivalent of
a classical average,
and there is no  interference apparent between states
that would signify their superposition or simultaneous occupancy.

The analysis was in terms of wave packets
situated on a coarse grid in the phase space
of the momenta and positions of the particles of the system.
The rationale for choosing these as the basis functions
lies in the observation that the state of a classical system
is completely characterized by the momenta and positions
of its constituent particles.
The trick in the analysis lies in reconciling this with
Heisenberg's uncertainty principle,
which is accomplished by choosing a sufficiently coarse grid
for the momentum and position labels of the wave packets.
Such a coarse grid serves to define the resolution
of the classical observation or measurement.
Provided that the classical phase functions are slowly varying
on the scale of the grid spacing,
then these momenta and positions  may be regarded as a continuum
for the purposes of classical mechanics.

Classical probability theory played an essential role in the analysis.
Superposition was defined to mean the simultaneous occupancy of states,
consequent upon which is their mutual interference.
In contrast classical probability theory is predicated
upon the exclusive occupancy of states:
the system must be in one, and only one, state at a time.\cite{TDSM}
The distinction between these two cases is reflected
in the diagonal and off-diagonal terms in the expectation values.
The presence or absence of the latter signifies
whether the system is quantum or classical.
An expectation value is a prediction
about the outcome of an observation or  measurement.
The case that only diagonal terms contribute to an expectation value
is a statement that the possible states of the system
are those in which exclusive occupancy holds.

\subsubsection{Further Questions}

Three further questions may be canvassed.
First, are these results consistent
with the existence of  macroscopic quantum phenomena?
Second, is the present mechanism the only way for
quantum systems to appear classical?
And third, are classical states superposed when nobody is looking?

It is evident that macroscopic quantum phenomena exist,
for example superfluidity and superconductivity.
The present results address
the dearth of classically observed superposition states.
They do not prohibit macroscopic quantum phenomena \emph{per se}.
Instead the present analysis says that
either  macroscopic quantum phenomena
are not formed from the superposition of classical states,
or else classical wave packets do not adequately represent
the quantum states relevant for macroscopic quantum phenomena.
Both statements are consistent with the present conclusion
that when a measurement or observation has limited resolution
(ie.\ projects onto classical space
defined by the span of the set of appropriately spaced wave packets),
then  a superposition of measured or observed states is precluded.

The origin of classicality does not solely reside
in the resolution with which observations or measurements are made.
The present definition of superposition
makes it clear that the absence or cancelation
of the off-diagonal terms in the expectation value
is essential for a system to be classical.
The alternative way that this can occur is if the wave function is decoherent,
since, by definition, coherency is required for interference effects.
In fact this is the same  problem as that of quantum measurement,
which related  wave function collapse
has been attributed to the  decoherence induced
by interactions with the environment.
One can mention the early work of Bohm,\cite{Bohm52}
and the more recent decoherence program
initiated by Zeh, Zurek, and Joos,\cite{Zeh70,Zurek82,Joos85}
the broad progress of which has been reviewed.
\cite{Zurek03,Schlosshauer04,Schlosshauer07,Galvan10,Kastner14}
The decoherence program holds that it is the interactions between a system
and its environment that causes the wave function to collapse
into the pointer states or the preferred basis of the measuring apparatus.
The former are said to be stable over time,
and the latter are said to preclude the superposition of states,
although different authors use different definitions.
\cite{Galvan10,Kastner14,Zurek81,Zurek93,Tegmark94,Wiseman98,Paz99,Busse09}
As one example,
Schlosshauer asks
``What singles out the preferred physical quantities in nature
---eg.\ why are physical systems usually observed to be in definite positions
rather than in superpositions of positions?'' (p.~50).
\cite{Schlosshauer07}
The preferred basis problem also occurs
in the Many Worlds Interpretation of quantum mechanics.\cite{Galvan10}
The precise relationship between decoherence and pointer states in measurement
on the  one hand,
and the emergence of classical behavior and its preferred basis on the other,
remains a matter of debate.
\cite{Zurek03,Schlosshauer04,Schlosshauer07,Galvan10,Kastner14,Zurek93}

In related work, the present author has shown
how exchange and entanglement with a thermodynamic reservoir
makes the sub-system wave function decoherent,
and the density operator diagonal in the entropy basis
(see appendix \ref{Sec:decohere}).\cite{QSM,STD2,Attard17}
For an equilibrium system,
the Maxwell-Boltzmann weighted distribution of entropy states
is sharply peaked and
remains stable during the time evolution of the open sub-system
(see the author's quantum\cite{QSM} and classical\cite{NETDSM}
analysis of the stochastic evolution of open sub-systems).
These two conclusions are a subtle modification of the usual definition
that the pointer states are individually stable
under time evolution.\cite{Galvan10}
The author formulates quantum statistical mechanics
as a formally exact integral over classical phase space,
with commutation function $W$ and symmetrization function $\eta$
formally accounting for the quantum corrections to classical behavior,
\begin{eqnarray} \label{Eq:<A>}
\left<  A(\hat{\bf p},\hat{\bf q}) \right>_{z,T}^\pm
& = &
\frac{1}{\Xi^\pm} \sum_{N=0}^\infty \frac{z^N}{h^{dN} N!}
\int \mathrm{d} {\bf p} \, \mathrm{d} {\bf q} \;
e^{-\beta {\cal H}({\bf p},{\bf q})}
\nonumber \\ & & \mbox{ } \times
A({\bf p},{\bf q}) W_A({\bf p},{\bf q}) \eta^\pm({\bf p},{\bf q}) .
\end{eqnarray}
The derivation of this result,
which is summarized in appendix \ref{Sec:qp},
uses exact position and momentum eigenfunctions, not wave packets.
The approach shows quantitatively how classical statistical mechanics
emerges at high temperatures and low densities,
in which limits the quantum corrections vanish,
$W_A({\bf p},{\bf q})  \rightarrow 1$
and
$\eta^\pm({\bf p},{\bf q})  \rightarrow 1$.
The significance of this result in the present context
is that it shows that decoherence is necessary
but not sufficient for classical behavior.

Finally,
the result of the present analysis
---that the predicted outcome of a measurement with classical resolution
precludes superposition states---
raises the question of whether or not a classical system
can exist in a superposition of states when nobody is looking
(ie.\ prior to measurement).
There are two remarks that can be made in response to this.

First,  the classical evolution equations
do not permit the bifurcation
of an initially non-superposed classical system,
and such equations are time reversible.
Assuming that the act of measurement or observation
is a negligible perturbation,
if the system is only ever observed  in one classical state
at a time, then it must always be in one classical state at a time,
even when it is not observed.

Second, since the only information we have about the world comes
from measurement,
and since one cannot take seriously a scientific theory
that is not falsifiable,
it is pointless and meaningless to speculate about
that which cannot be measured or observed.

Alternative to this second remark,
it is permissable to develop theories about regimes
that are beyond measurement,
but such theories should be judged by aesthetics, including simplicity,
and by their consequences when extrapolated to the measurable world.
Evidently the simplest view consistent
with the non-superposition of classical states upon measurement or observation,
is that classical states are not superposed
prior to measurement or when nobody is looking.

Simplicity as a criterion  for acceptable scientific theories
is attributed to William of Occam (1287--1347).\cite{Occam}
The so-called Law of Parsimony is often rendered as
``Entities are not to be multiplied without necessity''.
This could well be a slogan for the present paper.
In this spirit the present author is of the opinion
that it would be of some benefit
to apply Occam's razor to Schr\"odinger's cat.





\appendix

\section{Decoherence} \label{Sec:decohere}


This analysis  summarizes that in \S 7.2 and \S 7.3 of Ref.~\onlinecite{STD2}.

We seek the average of an operator on  an open sub-system.
To this end we begin with the  weight of states
of an isolated sub-system\cite{QSM,STD2}
\begin{equation}
\omega_{nn} = 1,
\mbox{ or }
\hat \omega = {\mathrm I}.
\end{equation}
This follows from Schr\"odinger's equation
and the axiom that time is uniform.

Label the non-degenerate energy macrostates by a Greek letter,
and the degenerate microstates by a Roman letter,
\begin{equation}
\hat{\cal H}^\mathrm{s} |  \zeta^\mathrm{Es}_{\alpha g} \rangle
  = E_\alpha^\mathrm{s} |  \zeta^\mathrm{Es}_{\alpha g} \rangle .
\end{equation}
In view of the preceding result
that microstates of an isolated sub-system have unit weight,
the entropy of an energy state is the logarithm of the number
of degenerate states that it contains,
$S_\alpha^\mathrm{E} = k_\mathrm{B} \ln N_\alpha^\mathrm{E}$,
where $N_\alpha^\mathrm{E} = \sum_{g \in \alpha} $.

For an open sub-system that can exchange  energy with a reservoir,
the total energy is
$E^\mathrm{tot} = E^\mathrm{s} + E^\mathrm{r} $.
The reservoir entropy for a particular energy is just
\begin{equation}
S^\mathrm{r}(E_\alpha^\mathrm{r}) =
k_\mathrm{B} \ln N_\alpha^\mathrm{Er},
\mbox{ with }
 N_\alpha^\mathrm{Er}
= \sum_{g\in \alpha} \!^{(\mathrm{Er})} .
\end{equation}
One also has the usual thermodynamic formula\cite{TDSM}
\begin{equation}
S^\mathrm{r}(E^\mathrm{r}) =
S^\mathrm{r}(E^\mathrm{tot}-E^\mathrm{s})
=
\mbox{ const. }
-\frac{E^\mathrm{s}}{T^\mathrm{r}}.
\end{equation}
The constant can be discarded,
as can the superscript on the reservoir temperature.


With $\{ | \zeta^\mathrm{Es}_{\alpha g} \rangle \}$
an orthonormal basis for the sub-system
and $\{ | \zeta^\mathrm{Er}_{\beta h} \rangle \}$
an orthonormal basis for the reservoir,
the most general wave function of the total system is
\begin{eqnarray}
| \psi_\mathrm{tot} \rangle
& = &
\sum_{\alpha,g; \beta,h} c_{\alpha g,\beta h}
| \zeta^\mathrm{Es}_{\alpha g}, \zeta^\mathrm{Er}_{\beta h} \rangle
\nonumber \\ & = &
\sum_{\alpha,g;h} c_{\alpha g , h}
| \zeta^\mathrm{Es}_{\alpha g} , \zeta^\mathrm{Er}_{\beta_\alpha h} \rangle .
\end{eqnarray}
Due to energy conservation,
the wave function is entangled,
which means that the expansion coefficient cannot be dyadic.\cite{Messiah61}
In the second equality,
the reservoir macrostate is defined implicitly,
$ E_\alpha^\mathrm{s} + E_{\beta_\alpha}^\mathrm{r} = E_\mathrm{tot} $.

With this, the expectation value of a sub-system operator is
\begin{eqnarray} \label{Eq:O(psi-tot)}
\lefteqn{
\left<\psi_\mathrm{tot} \right|
\hat A^\mathrm{s}
\left|\psi_\mathrm{tot} \right>
} \nonumber \\
& = &
\sum_{\alpha'g'; h'}
\sum_{\alpha g; h}
c_{\alpha' g' , h'}^*\,
c_{\alpha g , h}
\left< \zeta^\mathrm{Es}_{\alpha' g'},
\zeta^\mathrm{Er}_{\beta_{\alpha'}h'} \right|
\hat A^\mathrm{s}
\left| \zeta^\mathrm{Es}_{\alpha g},
\zeta^\mathrm{Er}_{\beta_{\alpha} h} \right>
\nonumber \\ & = &
\sum_{\alpha'g'; h'}
\sum_{\alpha g; h}
c_{\alpha' g' , h'}^*\,
c_{\alpha g , h}
\left< \zeta^\mathrm{Es}_{\alpha' g'}  \right|
\hat A^\mathrm{s}
\left| \zeta^\mathrm{Es}_{\alpha g} \right>
\left< \zeta^\mathrm{Er}_{\beta_{\alpha'}h'} \right|
\left. \zeta^\mathrm{Er}_{\beta_{\alpha} h} \right>
\nonumber \\ & = &
\sum_{\alpha,g,g'}  \sum_{h\in\beta_\alpha}\!\!^\mathrm{(Er)}\;
c_{\alpha g' , h}^*\,
c_{\alpha g , h}
\left< \zeta^\mathrm{Es}_{\alpha g'}  \right|
\hat A^\mathrm{s}
\left| \zeta^\mathrm{Es}_{\alpha g} \right>
\nonumber \\ & = &
\sum_{\alpha,g,g'}   \sum_{h\in\beta_\alpha}\!\!^\mathrm{(Er)}\;
c_{\alpha g' , h}^*\,
c_{\alpha g , h}\,
A^\mathrm{s,E}_{\alpha g',\alpha g} .
\end{eqnarray}
The third equality follows because the relationship between
$\alpha$ and $\beta_\alpha$ is bijective
(the macrostates are non-degenerate),
$\langle \zeta^\mathrm{Er}_{\beta_{\alpha'}h'} |
 \zeta^\mathrm{Er}_{\beta_{\alpha} h} \rangle
= \delta_{\alpha',\alpha} \, \delta_{h',h}$.
At this stage the density matrix in the energy (ie.\ entropy) representation
consists of principal blocks on the diagonal.

Since the microstates of the total system have equal weight,
apart from overall normalization
the coefficients
of the total wave function all have unit magnitude,
\begin{equation} \label{Eq:c-cond}
c_{\alpha g, h }
= e^{ i \theta_{\alpha g, h }} .
\end{equation}
The phases are uniformly and randomly distributed,
and so averaging over them gives
\begin{equation}
\left<
  c_{\alpha g',h}^* \,  c_{\alpha g,h} \,
\right>_\mathrm{stat}
=
\left<
e^{ i [\theta_{\alpha g, h }-\theta_{\alpha g', h }] }
\right>_\mathrm{stat}
=
\delta_{g,g'} .
\end{equation}
Hence
\begin{eqnarray} \label{Eq:delta_{g,g'}}
\left<
\sum_{h\in \beta_{\alpha}}\!^{(\mathrm{Er})}\;
\; c_{\alpha g',h}^* \,  c_{\alpha g,h} \,
\right>_\mathrm{stat}
& = &
\delta_{g,g'}
\sum_{h\in \beta_{\alpha}}\!^{(\mathrm{Er})}
\nonumber \\ & = &
\delta_{g,g'}\,
e^{ -E^\mathrm{s}_\alpha / k_\mathrm{B}T} ,
%
\end{eqnarray}
since
$k_\mathrm{B} \ln  N^\mathrm{Er}_{\beta_\alpha}
= S_{\beta_\alpha}^\mathrm{r}
=  -E^\mathrm{s}_\alpha /T$.
With this the expectation value of the sub-system operator
in the total system is
\begin{equation} \label{Eq:RPA}
\left< \hat A^\mathrm{s} \right>_\mathrm{stat}
=
\frac{1}{Z}  \sum_{\alpha,g}
 e^{ -\beta E_\alpha }
A^\mathrm{s,E}_{\alpha g,\alpha g}
=
\mbox{Tr } \hat \wp \hat A^\mathrm{s} ,
\end{equation}
where the probability operator is
$ \hat \wp  = Z^{-1} e^{\beta \hat {\cal H}  }$
and $\beta \equiv 1/ k_\mathrm{B}T$.
One sees that there is no superposition of states in this
averaged expectation value;
the wave function of the sub-system has collapsed into energy (entropy)
microstates.
Since the variables that are not exchanged with the reservoir
(eg.\ number in this case)
belong to the degenerate states,
the states corresponding to these have also collapsed.

Its worthwhile showing that the random phase approach
gives the same result as an integral over the total wave space,
namely
\begin{eqnarray} \label{Eq:O-stat-vN}
\lefteqn{
\left< \hat A^\mathrm{s} \right>_\mathrm{stat}
} \nonumber \\
& = &
\frac{1}{Z'} \int \mathrm{d} \psi_\mathrm{tot} \;
A^\mathrm{s}(\psi_\mathrm{tot})
\nonumber \\ & = &
\frac{1}{Z'}
\sum_{\alpha,g,g'}
\sum_{h\in\beta_\alpha}\!\!^\mathrm{(Er)}\;
A^\mathrm{s,E}_{\alpha g',\alpha g}
\int \mathrm{d} \underline{\underline c} \;
\frac{ c_{\alpha g',h}^* \, c_{\alpha g, h} }{ N(\psi_\mathrm{tot}) }
\nonumber \\ & = &
\frac{1}{Z'} \sum_{\alpha,g}
\sum_{h\in\beta_\alpha}\!\!^\mathrm{(Er)}\;
A^\mathrm{s,E}_{\alpha g,\alpha g}
\int \mathrm{d} \underline{\underline c} \;
\frac{ | c_{\alpha g, h}|^2 }{ N(\psi_\mathrm{tot}) }
\nonumber \\ & = &
\frac{1}{Z'}
\sum_{\alpha,g}
e^{ S_{\beta_\alpha}^\mathrm{r} / k_\mathrm{B}}
A^\mathrm{s,E}_{\alpha g,\alpha g} \times \mbox{const.}
\nonumber \\ & = &
\frac{1}{Z}  \sum_{\alpha,g}
 e^{ -E_\alpha^\mathrm{s} / k_\mathrm{B}T}
A^\mathrm{s,E}_{\alpha g,\alpha g} .
\end{eqnarray}
The third equality follows since the terms
in which the integrand  is odd  vanish;
the only non-vanishing terms have $g=g'$.
The fourth equality follows because all the integrations
give the same constant independent of $\alpha$, $g$, and $h$.

This analysis shows that that the  statistical average
of a sub-system quantum operator
is just the von Neumann trace,
with the density operator being
the Maxwell-Boltzmann (or Gibbs more generally) operator,
which is diagonal in the entropy basis.
For the present,
the important point is that this mixture formulation
results directly from the decoherence induced in the wave function
by exchange, entanglement, and subsequent averaging
over the reservoir states.
The time scale for entanglement and decoherence
is likely much shorter than that for equilibration.
\cite{Brune96,Hackermuller04}

%
\section{Phase Space Representation} \label{Sec:qp}
%

This analysis summarizes and improves
that given in \S 7.4 of Ref.~\onlinecite{STD2}.

The grand partition function
for a sub-system that is entangled with a reservoir
by the exchange of energy and number  is\cite{QSM,STD2}
\begin{eqnarray}
\Xi^\pm
& = &
\sum_{N=0}^\infty \frac{z^N}{N!} \sum_{\hat P} (\pm 1)^p \sum_{\bf n}
\left\langle \hat P {\bf n} \left|
e^{-\beta \hat {\cal H} }
\right| {\bf n} \right\rangle
\nonumber \\ & = &
\mbox{TR}'\, e^{-\beta \hat {\cal H} }.
\end{eqnarray}
Here $z$ is the fugacity, $N$ is the number of particles,
$\beta$ is the inverse temperature,
and $|{\bf n}\rangle$ is a complete set of energy eigenfunctions.
Also, $\hat P $ is the permutation operator,
$p$ is its parity, the upper sign is for bosons
and the lower sign is for fermions.
This has the form of a trace over distinct states.
Similarly the statistical average of an operator is
\begin{equation}
\left< \hat A \right>_{z,\beta}^\pm
=
\frac{1}{\Xi^\pm} \mbox{TR}'\, e^{-\beta \hat {\cal H} } \hat A.
\end{equation}
As these are in the form of a trace, they can be written in any basis.

The position representation for $N$ particles in $d$ dimensions is
${\bf r} = \{ {\bf r}_1, {\bf r}_2, \ldots, {\bf r}_N\}$,
with ${\bf r}_j = \{ r_{jx}, r_{jy}, \ldots, r_{jd}\}$.
The orthonormal momentum eigenfunctions
in the position representation are\cite{Messiah61,STD2}
\begin{equation}
|{\bf p}\rangle
\equiv
\frac{e^{-{\bf p}\cdot{\bf r}/i \hbar}}{V^{N/2}} ,
\end{equation}
where $V=L^d$ is the volume of the sub-system.
The momentum label ${\bf p}$ is discrete
with the spacing between momentum states being\cite{Messiah61}
$\Delta_p = 2\pi \hbar /L$.
The position eigenfunctions
in the position representation
are just the Dirac-$\delta$ functions,\cite{Messiah61}
\begin{equation}
|{\bf q}\rangle
\equiv
\delta({\bf r}-{\bf q})  .
\end{equation}
The momentum and position eigenfunctions each form a complete set.
Hence the partition function can be re-written
\begin{eqnarray}
\Xi^\pm
& = &
\sum_{N=0}^\infty \frac{z^N}{N!} \sum_{\hat P} (\pm 1)^p \sum_{\bf p}
\left\langle \hat P {\bf p} \left|
e^{-\beta \hat {\cal H} }
\right| {\bf p} \right\rangle
\nonumber \\ & = &
\sum_{N=0}^\infty \frac{z^N}{N!}  \Delta_p^{-dN} \sum_{\hat P} (\pm 1)^p
\int \mathrm{d}{\bf p}\; \mathrm{d}{\bf q}\;
\nonumber \\ &  & \mbox{ } \times
\left. \left\langle \hat P {\bf p}  \right| {\bf q} \right\rangle
\; \left\langle  {\bf q} \left|
e^{-\beta \hat {\cal H} }
\right| {\bf p} \right\rangle
\nonumber \\ & = &
\sum_{N=0}^\infty \frac{z^N}{h^{dN} N!}
\sum_{\hat P} (\pm 1)^p
\int \mathrm{d}{\bf q} \, \mathrm{d}{\bf p}\;
\nonumber \\ && \mbox{ } \times
\frac{  \left\langle  {\bf q} \left|
e^{-\beta \hat {\cal H} }
\right| {\bf p} \right\rangle
}{\langle  {\bf q} | {\bf p} \rangle }
\frac{ \left. \left\langle \hat P {\bf p}  \right| {\bf q} \right\rangle
}{\langle  {\bf p} | {\bf q} \rangle }
\nonumber \\ & \equiv &
\sum_{N=0}^\infty \frac{z^N}{h^{dN}N!}
\int\!\! \mathrm{d}{\bf \Gamma}
 e^{-\beta {\cal H}({\bf \Gamma})}\,
 W({\bf \Gamma})\,
 \eta^\pm({\bf \Gamma}) .
\end{eqnarray}
Here a point in classical phase  has been denoted
${\bf \Gamma} \equiv \{{\bf p} , {\bf q} \}$.
The commutation function $W$ is defined via
\cite{Wigner32,Kirkwood33,STD2,Attard17}
\begin{equation}
 e^{-\beta {\cal H}({\bf \Gamma})}\,
 W({\bf \Gamma})
 \equiv
\frac{  \left\langle  {\bf q} \left|
e^{-\beta \hat {\cal H} }
\right| {\bf p} \right\rangle
}{\langle  {\bf q} | {\bf p} \rangle } ,
\end{equation}
and the symmetrization function is
\cite{QSM,STD2,Attard17}
\begin{equation}
 \eta^\pm({\bf \Gamma})
 \equiv
\sum_{\hat P} (\pm 1)^p
\frac{ \left. \left\langle \hat P {\bf p}  \right| {\bf q} \right\rangle
}{\langle  {\bf p} | {\bf q} \rangle } .
\end{equation}
An infinite resummation of the symmetrization function
casts it as a series of loop permutations,
which is quite tractable computationally.\cite{QSM,STD2,Attard17}
The commutation function and the symmetrization function are complex.
But the imaginary parts of each are odd in momentum,
and so the grand partition function is real.

Analogous analysis for the statistical average yields
the same result with
$e^{-\beta \hat {\cal H} } \Rightarrow e^{-\beta \hat {\cal H} } \hat A$,
\begin{eqnarray}
\left< \hat A \right>_{z,\beta}^\pm
& = &
\frac{1}{ \Xi^\pm }
\sum_{N=0}^\infty \frac{z^N}{h^{dN}N!}
\int\!\! \mathrm{d}{\bf \Gamma}
\nonumber \\ && \mbox{ } \times
 e^{-\beta {\cal H}({\bf \Gamma})} \,  A({\bf \Gamma})\,
 W_A({\bf \Gamma})  \,
 \eta^\pm({\bf \Gamma}) .
\end{eqnarray}
Here $\hat A = A(\hat{\bf p},\hat{\bf q})$, and
\begin{equation}
 e^{-\beta {\cal H}({\bf \Gamma})}\,
 A({\bf \Gamma})
 W_A({\bf \Gamma})
 \equiv
\frac{  \left\langle  {\bf q} \left|
e^{-\beta \hat {\cal H} } A(\hat{\bf p},\hat{\bf q})
\right| {\bf p} \right\rangle
}{\langle  {\bf q} | {\bf p} \rangle } .
\end{equation}
For a Hermitian operator $A({\bf p},{\bf q})$ is real,
as must be the statistical average.
Since  $\eta^\pm({\bf p},{\bf q})^* = \eta^\pm(-{\bf p},{\bf q})$,
it is sufficient that
$ A({\bf p},{\bf q})  W_A({\bf p},{\bf q})^*
= A(-{\bf p},{\bf q})  W_A(-{\bf p},{\bf q})$,
in order for the imaginary part to integrate to zero.
If $ A(\hat{\bf p},\hat{\bf q}) =  A_p(\hat{\bf p}) + A_q(\hat{\bf q})$,
then it may be shown that one can make the replacement
$W_A({\bf \Gamma}) \Rightarrow W({\bf \Gamma})$.

It is worth mentioning that
the loop factorization of the symmetrization function
facilitates the evaluation of averages
that can be written as a derivative of the grand potential.

Since the average must be real,
the imaginary part must always integrate to zero
no matter how one formulates the integrand.
Nevertheless,
it may be more efficient computationally to write it
in the most symmetric fashion using the fact that the trace operation
is insensitive to the order of the operators
or of the position and momentum eigenfunctions.
Hence one can use the symmetric integrand
\begin{eqnarray}
\lefteqn{
 e^{-\beta {\cal H}({\bf \Gamma})}\,
 A({\bf \Gamma})
\overline W_A({\bf \Gamma})
\overline \eta^\pm({\bf \Gamma})
} \nonumber  \\
&  \equiv &
\frac{1}{4} \left\{
\frac{  \left\langle  {\bf q} \left|
e^{-\beta \hat {\cal H} } \hat A
+
 \hat A  e^{-\beta \hat {\cal H} }
\right| {\bf p} \right\rangle
}{\langle  {\bf q} | {\bf p} \rangle } \eta^\pm({\bf \Gamma})
\right. \nonumber \\  &  & \left. \mbox{ }
+
\frac{  \left\langle  {\bf p} \left|
e^{-\beta \hat {\cal H} } \hat A
+
 \hat A  e^{-\beta \hat {\cal H} }
\right| {\bf q} \right\rangle
}{\langle  {\bf p} | {\bf q} \rangle } \eta^\pm({\bf \Gamma})^*
\right\}
\nonumber \\ & = &
\frac{1}{2} \mbox{Re} \left\{
\frac{  \left\langle  {\bf q} \left|
e^{-\beta \hat {\cal H} } \hat A
+
 \hat A  e^{-\beta \hat {\cal H} }
\right| {\bf p} \right\rangle
}{\langle  {\bf q} | {\bf p} \rangle } \eta^\pm({\bf \Gamma})
\right\} .
\end{eqnarray}

Finally, at high temperatures,
\begin{equation}
W({\bf \Gamma}) \rightarrow 1 , \;\; \beta \rightarrow 0,
\end{equation}
and at high temperatures or low densities,
\begin{equation}
\eta^\pm({\bf \Gamma}) \rightarrow 1
, \;\; \beta \rightarrow 0 \mbox{ or } z \rightarrow 0.
\end{equation}
In these limits the partition function and statistical average
become purely classical.

Because $W_A$ and $\eta^\pm$ are complex,
the phase space integral does not generally take on the form
of a classical average,
since the weights in the latter must be real and non-negative.\cite{TDSM}
As just mentioned, the integrand can be written in a symmetric fashion
that is real, but there is no guarantee that it is non-negative.
Also, in the general case the commutation function $W_A$ depends
upon the phase function being averaged,
whereas a classical weight density should depend only on the state
independent of the function of the state being averaged.
The functions $W_A$ and $\eta$ only become fully classical
in the high temperature, low density regime.
So although decoherence from the entanglement with the reservoir
leads directly to the suppression of the superposition
of both energy states and  phase (ie.\ momentum and position) states,
decoherence alone is not sufficient to ensure full classical behavior.



\end{document}